\pdfoutput=1
\documentclass[11pt]{article}
\usepackage[final]{acl}
\usepackage{times}
\usepackage{latexsym}
\usepackage{stfloats}
\usepackage{multirow}
\usepackage[title]{appendix}
\usepackage{xfakebold}
\usepackage{graphicx}
\usepackage{natbib}
\usepackage{caption}
\usepackage{float}
\usepackage{tabulary,booktabs}
\usepackage{subcaption}
\usepackage{amsmath}
\usepackage[T1]{fontenc}
\usepackage[utf8]{inputenc}
\usepackage{microtype}
\usepackage{inconsolata}

\title{Forecasting Credit Ratings: A Case Study where Traditional Methods Outperform Generative LLMs}

\author{Felix Drinkall*, Janet B. Pierrehumbert*\ddag, Stefan Zohren*\dag \\
    *Department of Engineering Science, University of Oxford \\
    \dag The Alan Turing Institute \\
    \ddag Faculty of Linguistics, University of Oxford \\
    \texttt{felix.drinkall@eng.ox.ac.uk}}

\begin{document}
\maketitle
\begin{abstract}
  Large Language Models (LLMs) have been shown to perform well for many downstream tasks. Transfer learning can enable LLMs to acquire skills that were not targeted during pre-training. In financial contexts, LLMs can sometimes beat well-established benchmarks. This paper investigates how well LLMs perform at forecasting corporate credit ratings. We show that while LLMs are very good at encoding textual information, traditional methods are still very competitive when it comes to encoding numeric and multimodal data. For our task, current LLMs perform worse than a more traditional XGBoost architecture that combines fundamental and macroeconomic data with high-density text-based embedding features. We investigate the degree to which the text encoding methodology affects performance and interpretability. The dataset reconstruction and model code from this paper is provided\footnote{https://github.com/FelixDrinkall/credit-ratings-project}.
\end{abstract}
\section{Introduction}

Corporate credit ratings indicate a borrower's ability to service its debt obligations and are a forward-looking measure of a company's health \cite{baresa2012role}. A company's credit rating is significant since it affects the cost of raising capital, which in turn could finance future infrastructure to increase revenue or profitability. An optimistic rating can result in a virtuous cycle whereby it is easier to raise money and grow the business \cite{optimism_cr}, and a pessimistic rating can result in a vicious cycle in which competition can grow faster due to cheaper debt obligations. Knowing which cycle a company may enter can be advantageous to investors. Many major funds are also not allowed to own sub-prime assets, which makes forecasting a drop in credit rating very important so that the fund has more time to divest from the asset, which could result in a higher close price. 

\begin{figure}
    \centering
    \includegraphics[width=\linewidth]{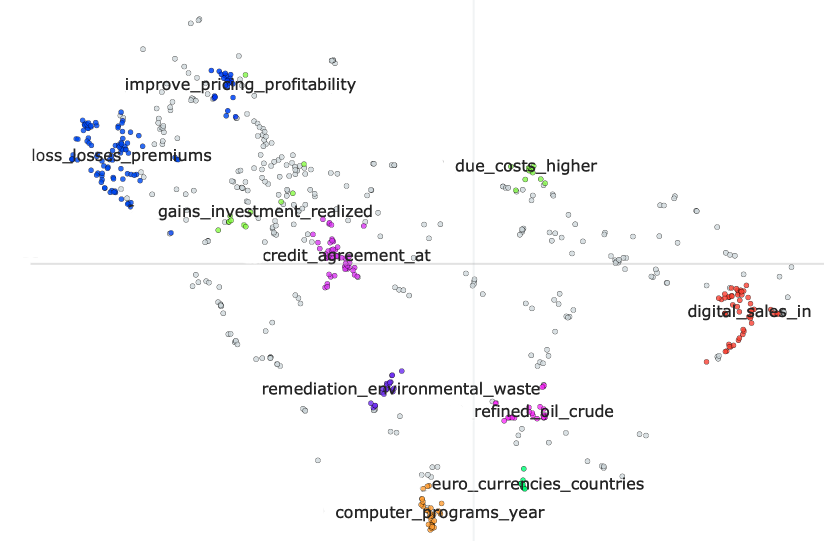}
    \caption{Example of the best-performing feature - high-density clustering \cite{drinkall-etal-2022-forecasting}. Each dot represents a sentence, and the colored areas representing high-density regions of the embedding space.}
    \label{fig:clusters}
\end{figure}

Recently, there has been a surge of interest in text-based forecasting \cite{xu-cohen-2018-stock, yang2020html, nie2024surveylargelanguagemodels}. One reason for this trend is the progress that has been made in text modelling in general \citep{wei2022emergent, touvron2023llama}. Given that financial news is often first disseminated through written or spoken communications \cite{10.1093/rof/rfw018}, rather than in numeric or tabular formats, there has been a hope that important information can be included in models sooner than was possible without using linguistic information. Another reason is that language can provide relevant context and forward-looking information, whereas financial numeric reporting alone is inherently retrospective. Contained within a company's filings, text can provide insights about the future strategic direction of the company as well as historical information.

The majority of text-based forecasting research has been focused on short text sequences, drawing primarily from sources such as social media \cite{xu-cohen-2018-stock}, news articles \cite{Zhang_2018}, and analyst recommendations \cite{rekabsaz-etal-2017-volatility}. In contrast, many fundamental financial documents, such as company filings, earnings call transcripts, and patents, are very long. Considering that the shorter texts often serve as summaries, reflections or commentaries on the detailed primary sources and that speed of information acquisition is essential in finance \cite{RZAYEV2023100853}, there must be more focus on text-based forecasting for longer text sequences. This paper evaluates the most effective ways to model longer text sequences within a text-based forecasting task.

Linked to the recent progress of LLMs, there has been growing interest in applying these models in a variety of downstream applications \cite{kaddour2023challenges}. It has been shown that as generative LLMs scale, they acquire abilities that were not present in smaller LLM variants \cite{wei2022emergent}, such as modular arithmetic \cite{srivastava2023imitation}, NLU \cite{hendrycks2021measuring}, commonsense reasoning \cite{lin2022truthfulqa}, fact-checking \cite{rae2022scaling} and so on. These abilities have been impressive, but it is best not to be over-optimistic.  The lack of training data transparency associated with some of the best-performing LLMs means that we cannot be certain whether some of the performance gains are due to the memorisation of benchmarks being in the training datasets \cite{stoch_parrots, sainz-etal-2023-nlp, xu2024benchmarking, balloccu-etal-2024-leak}. Generative LLMs also seem to have a mediocre understanding of concepts like negation and complex logical reasoning \cite{kassner-schutze-2020-negated, lorge-pierrehumbert-2023-wacky, huang-chang-2023-towards, truong-etal-2023-language}. 
These limitations in the capabilities of LLMs could prove to be very consequential in financial contexts.  In this paper, we test generative LLMs on a complex linguistic task, which has never been fully solved by human experts: credit rating forecasting. We show that while LLMs encode text-based information very well, they are not good at incorporating numeric information, and underperform a boosting-tree baseline.

The contributions of this paper are as follows:

\begin{itemize}    
    \item We show that generative LLMs are poor at encoding numerical information, and underperform traditional methods.
    \item To our knowledge, this is the first use of modern language modelling techniques in a credit rating forecasting task.
    \item A financial dataset that can be reproduced with an academic WRDS licence.
    \item A benchmark of techniques for encoding long-sequence text in a forecasting task.
\end{itemize}
\section{Related Work}

\subsection{Text-based forecasting}
\label{sec:tbf_main}

\subsubsection{Encoding Text for Forecasting}
\label{sec:tbf_feat}

The predominant approach in text-based forecasting has focused on the extraction of interpretable features like sentiment and uncertainty scores \cite{song2019forecasting, an2023text}. Rule-based sentiment using diverse lexicons has dominated the literature \cite{mohammad2020practical, kalamara2022making, barbaglia2023forecasting}. Lexicons tailored to specific domains generally surpass broader lexicons in predictive tasks \cite{loughran2011liability, li2014news}. Nevertheless, lexicons overlook contextual nuance and inadequately address common linguistic phenomena like negation. To mitigate these limitations, efforts have been made to integrate more sophisticated sentiment classifiers \cite{an2023text, 10134443}. However, sentiment presupposes that important information can be encapsulated within a single dimension. To avoid an overly simple and prescriptive feature set, unsupervised methods have been used in feature exploration: TF-IDF \cite{jones1972statistical}, Latent Dirichlet Allocation (LDA) \cite{wang2017study, Kanungsukkasem2019FinancialLD}. However, the arrival of contemporary topic models has gradually eclipsed LDA, fostering the adoption of transformer-derived topic models into forecasting tasks \cite{drinkall-etal-2022-forecasting}. 

Recently, some studies have used the representations from encoder-based LLMs as features for text-based forecasting. LLMs exploit high-dimensional embeddings to capture the linguistic meaning of words \cite{devlin-etal-2019-bert, radford2019language} and sentences \cite{reimers-gurevych-2019-sentence}, with representation dimensionality ranging from 384 \cite{wang2020minilm} to 5192 \cite{touvron2023llama}. Such dimensionality poses challenges when used with smaller datasets. Nonetheless, there has been some success in incorporating these methods into text-based forecasting \cite{sawhney-etal-2020-deep, lee2023moat}. However, the effectiveness of LLMs is often hampered by their limited context windows. Recent advancements have seen an increase in context window sizes, thanks in part to better GPU infrastructure making it computationally feasible, the implementation of attention sparsification techniques \cite{efficient_transformers}, and positional encoding hacks \cite{chen2023extending}. There are also methods that combine the use of transformer-based LLMs with feature-based methods, such as topic clusters \cite{grootendorst2022bertopic, drinkall-etal-2022-forecasting}, or emotions \cite{liapis2023temporal}.

\subsubsection{Generative Multimodal Forecasting}

In addition to the adoption of encoder-based LLMs like BERT \cite{devlin-etal-2019-bert}, generative LLMs have been used in text-based forecasting tasks. Generative LLMs use masked-self attention to model text in an autoregressive manner. The GPT \cite{radford2019language} and Llama \cite{touvron2023llama} model families are part of the generative model class. LLMs have been used as a backbone model for generative time-series forecasting models \cite{cao2023tempo, chang2023llm4ts, zhou2023fits, liu2024unitime}, showing that an adapted generative language model can forecast weather, electricity and several other domains without relying on traditional text inputs. \citealt{liu2024unitime} used eight text-based frames in order to create a general time-series modal that could be applied to several domains, and in so doing encoded both text and numerical information in a GPT-2-small model to generate the predicted time-series.

Beyond the use of text-based frames in generative forecasting tasks, GPT4MTS \cite{lee2023moat} encoded both news and time-series information before passing the concatenated input sequence through a pre-trained GPT-2 model. FinMA \cite{xie2023pixiu} and PromptCast \cite{xue2023promptcast} evaluated the performance of LLMs on stock movement prediction by converting the time-series information into natural language and prompting the language model for the predicted direction. \citealt{yu2023temporal} takes this further by passing exclusively text information into the prompt for a financial forecasting task. There has been little comparison between these generative methods and the more traditional discriminative methods when applied to multimodal information.

\subsection{Credit Rating Prediction}

Research in \textbf{C}redit \textbf{R}ating \textbf{P}rediction (CRP) has tended to focus on predicting the absolute credit rating at time \(t=0\)  given the feature set \(F_{t=0}\) \cite{electronics12071643, GALIL2023104648, tavakoli2023multimodal}. This approach takes the perspective of the rating agencies and is useful for identifying anomalies where the existing credit rating classification appears to be implausible or inconsistent with current financial indicators \cite{lokanan2019detecting}. However, predicting the absolute rating level is more simple and not as useful as predicting a future change. There is some limited research on \textbf{C}redit \textbf{R}ating \textbf{F}orecasting (CRF), where the target is the movement direction of the credit rating at time \(t=1\). This task takes the perspective of the investor seeking to predict whether an asset is likely to be classified as more or less risky in the next time period, and is the task outlined in this paper.

There have been some attempts to incorporate linguistic information into both corporate risk \cite{FEI2015725, cao2024risklabs} and default prediction \cite{MAI2019743, STEVENSON2021758}. Some papers have shown how text can help improve consumer credit lending \cite{hurley2016credit, BABAEI2024100534}. There has also been some attempts to include textual data in CRP and CRF tasks \cite{10.1016/j.eswa.2022.118042, munoz2022machine, tavakoli2023multimodal}. The majority of the existing literature uses lexicons, keywords or sentiment to encode the text \cite{Kogan2009PredictingRF, FEI2015725, MAI2019743, munoz2022machine, 10.1016/j.eswa.2022.118042}. There have been some studies that have utilized encoder-based LLM representations \cite{STEVENSON2021758, tavakoli2023multimodal, cao2024risklabs}. There has been some work exploring how well generative models perform at assessing credit lending applications \cite{BABAEI2024100534}, and value at risk in general \cite{cao2024risklabs}, but there has been no work benchmarking how well modern generative LLMs perform on a CRF task. Understanding generative LLMs' relative strengths relative to more traditional methods is an important contribution to the existing literature.

\section{Dataset}

In part due to the lack of large open-source or readily available datasets with temporal metadata, most of the financial text-based forecasting studies have either focused on expensive proprietary datasets, or datasets spanning 2-3 years \cite{xu-cohen-2018-stock, soun2022accurate}, making results hard to replicate and potentially biased to a specific time. While temporal bias in language-based tasks is hard to avoid due to limited historical data \cite{drinkall-etal-2024-time}, we aim to reduce this by using a dataset spanning 23 years, increasing the models' exposure to different economic contexts. The cost and lack of transparency of large datasets have hindered progress in the field and made it harder to build on promising work due to the difficulty of replicating results. As such, all data used in this paper is either open source or available with a WRDS subscription to enable effective dataset reconstruction. The data used in this paper is from US-based companies. 

\subsubsection{Credit ratings (C)}

For the credit ratings, we used the Compustat Capital IQ dataset\footnote{Credit Ratings: https://tinyurl.com/r4urtkc5}, using Standard \& Poors' (S\&P) ratings. These ratings cover the period from 1978 to 2017. S\&P routinely assesses and assigns credit ratings to companies. Our paper predicts changes to the long-term credit ratings. Notably, we incorporate historical ratings from preceding quarters into our prediction models, acknowledging the distinct implications of a top-rated company (AAA) being downgraded compared to a lower-rated one (CC) experiencing a similar decline.

\subsubsection{SEC filings}

This paper uses 10-Q and 10-K filings available in the SEC's EDGAR database\footnote{Filings Database: https://tinyurl.com/3rdn7hrx} to provide both textual context. They were chosen for their consistent structure which aids homogenous feature extraction. While most of the content in these filings is comprised of indexing, tables, and introductory text, we're interested in the parts that offer insights into a company's future financial health. As such, we've focused on the Management's Discussion and Analysis of Financial Condition and Results of Operations (MDA) section. We extract the MDA sections from all SEC filings - using the SEC-API\footnote{Filings API: https://sec-api.io/} - for which we had credit rating data, spanning from Q1-1994 to Q2-2017. The API returns cleaned text, but we clean the text further by removing the remaining HTML, links and excessive spaces.

\subsubsection{Fundamental data (F)}
\label{sec:fund_data}

S\&P emphasizes two components in their credit rating methodology: the financial and business risk profiles \cite{gillmor2015}. While the text from the MDA section provides some insight into the qualitative business risk profile, numerical fundamental data is important to assess the financial health of a company. For the fundamental data, we use the Compustat Quarterly Fundamentals dataset\footnote{Fundamental Data:https://tinyurl.com/4ca8ddst}. The variables selected are outlined in Appendix \ref{app:var_desc}. These variables were consistently reported for all the companies under consideration. Ideally, we would incorporate a broader range of fundamental variables, but expanding the variable set would result in fewer samples with complete data, thus limiting the scope of our analysis.

\subsubsection{Macroeconomic data (M)}
\label{sec:macro_data}

Adverse events in the world economy can also impact a company's ability to repay its debt. Many external forces can affect a company's future creditworthiness, however, we have identified three key areas from prior research in the area \cite{carling2007corporate, TAYLOR2021102432}: labour statistics, interest rates and foreign exchange data. For the labour statistics, we used the Bureau of Labour Statistics dataset\footnote{Labour Statistics: https://tinyurl.com/y94d52xk}. For the interest rate and foreign exchange data, we used the Federal Reserve Bank Reports\footnote{Interest Rate Data: https://tinyurl.com/46aw6mu2}$^\text{,}$\footnote{Foreign Exchange Data: https://tinyurl.com/a38rmzd8}. 

\subsection{Dataset Construction}

To maintain consistent periodicity in SEC filings, all data is aligned quarterly. The dataset spans from Q1 1994, when the SEC began electronic processing of filings, to Q2 2017, the last period with credit rating data from Compustat Capital IQ. Companies with incomplete records were excluded. As a result, when the number of lagged quarters used in the task is increased, the number of valid samples diminishes. This reduction is due to the lower probability of having complete data across many consecutive quarters, compared to when only the most recent quarter is considered.

Credit rating data is highly imbalanced, with 93.4\% of companies maintaining the same score. While oversampling techniques like SMOTE \cite{chawla2002smote} are common for credit rating prediction \cite{ijfs11030096, WANG2022108153, Zhao2024ResamplingTS}, their application to text embeddings lacks consensus. To address this, we balanced the classes, reducing the dataset size. Training data spans Q1 1994 to Q4 2012, validation from Q1 2013 to Q4 2014, and testing from Q1 2015 to Q4 2016. The dataset was made from 23 years and the size is representative of many other tasks in NLP (Table \ref{tab:dataset_size}).

The MDA section of an SEC Filing, despite only constituting a small part of the filing, is still very long. The average MDA section in our task is 13,267 tokens long using a BPE tokenizer \cite{sennrich-etal-2016-neural}. As such, when a model was not able to encode all of the tokens, only the first part of the text was encoded.

\section{Methodology}
\label{sec:methodoolgy}

We deploy two frameworks to test different architectural methodologies on this task. The same data are provided to each of the frameworks. The first framework is a feature-based discriminative approach that uses a more traditional boosting-tree model and tests the different ways to encode the textual data. The second uses generative LLMs and prompting to output one of a fixed list of labels through a greedy search algorithm (App. \ref{app:greedy_decode}).

\subsection{Task Description}
\label{sec:task_desc}

The objective is to predict the credit rating, \( \hat{R}_t \), at time \( t \). The function can be represented as follows:

\[
\begin{aligned}
\hat{R}_t = f\big( &T_{t-1}, T_{t-2}, \ldots, T_{t-p}; \\
                  &R_{t-1}, R_{t-2}, \ldots, R_{t-p}; \\
                  &N_{t-1}, N_{t-2}, \ldots, N_{t-p} \big)
\end{aligned}
\]
Here, \( T_{t-i} \) represents the text data, \( R_{t-i} \) represents the historical credit rating data, \( N_{t-i} \) represents the numeric data - both fundamental and macroeconomic. \(i\) varies from 1 to \(p\), with \(p\) indicating the number of past quarters considered (1 to 4 quarters in this study). Furthermore, \( f \) is the predictive function to convert the input data into an estimate. An ablation study is conducted to evaluate the impact of different data types on the prediction accuracy. In this study, the function \(f\) is tested under various configurations: using only text data \( T_{t-i} \), using combinations of historical ratings \( R_{t-j} \), and numeric data \( N_{t-k} \). This approach helps to determine the relative importance of each type of data.

\subsection{Boosting-Tree Baseline}

To test the abilities of generative LLMs, it is necessary to benchmark the performance against a relatively well-understood and robust algorithm. We select XGBoost \cite{chen2016xgboost}, a model that has been widely adopted in many domains \cite{talukder2023dependable, DONG2023105579, joshi2024application}. The supervised model takes as input the normalized fundamental, macroeconomic and text data, and outputs the most likely label. We describe other more complex neural network architectures that failed to learn this task in Appendix \ref{app:hier_cr}. Due to the restricted dataset size, the models outlined in the Appendix were unable to learn the task before overfitting the training dataset. 

\subsection{Text Encoders}

To test and identify which of the traditional encoders performs best we trialled a series of standard methodologies.

The Loughran McDonald Lexicon (\textbf{LM}) \cite{loughran2011liability} is widely recognized in finance. Given its prevalence, it is crucial to compare its effectiveness with more advanced methods. The lexicon classifies words into four domains: Positivity, Negativity, Litigiousness, and Uncertainty. However, the simple language modelling technique classifies phrases like "The debt increased last quarter" as neutral. The LM text representation in this work is the document word count from each sentiment, normalized by the maximum value in the training set.

Latent Dirichlet Allocation (\textbf{LDA}) is a widely-used topic modeling method that identifies latent topics within text \cite{blei2003latent}. It operates by assuming that each document is a mixture of topics and that each topic is a distribution over words. Despite advancements in topic modeling, LDA remains a reliable baseline for evaluating newer models. In this paper, the features represent texts as probability distributions over 25 topics, with each dimension indicating the likelihood of the text belonging to a specific topic.

High-density Embedding Clusters (\textbf{HEC}) leverage the natural language understanding of LLMs but reduce the dimensionality of the input feature. HEC provides a good basis for topic modelling \cite{Sia2020, grootendorst2022bertopic}. Sentence embeddings have also been used to discern domain type from text \cite{Aharoni2020}. \citealt{drinkall-etal-2022-forecasting} extended this work to generate features from clusters of sentence embeddings in a COVID-19 caseload prediction task. For this task, each sentence of each filing in the training set was encoded into embeddings space using a \textit{all-mpnet-base-v2} \cite{reimers-gurevych-2019-sentence}, the dimensionality was then reduced using UMAP \cite{mcinnes2020umap}, and the HDBSCAN clustering algorithm \cite{campello2013density} was used for form 100 distinct clusters. An example of the cluster features is displayed in Figure \ref{fig:clusters}. 
Then each filing in the train, validation and test set was split into sentences and then transformed into the embedding space described above. The overall text representation was the average of the representations of each sentence, and the representation of each sentence was the probability distribution that the sentence belonged to each of the 100 clusters.

To understand the extent to which emotion scores play a significant role in forecasting the next credit rating. We used a DistilRoBERTa model ($\boldsymbol{E_{\text{DRoBERTa}}}$) that had been fine-tuned on an emotion classification task \cite{hartmann2022emotionenglish}. The SEC Filings are then chunked into 512 token sequences and classified according to the probability that that chunk can be associated with each emotion. The average across all chunks is taken as the final text representation of each filing.

We also trialled a pooled MP-NET representation \cite{song2020mpnetmaskedpermutedpretraining} by chunking the text into 512 token segments and averaging over the pooled representation of each chunk. In well-established benchmarks \cite{muennighoff2023mtebmassivetextembedding}, MP-NET embeddings have performed strongly for their size and provide a baseline comparison to the HEC features derived from the MP-NET model.



\subsection{Generative Framework}
\label{sec:gen_framework}

Given recent advancements in generative LLMs, we evaluate whether these models can identify changes in a company's perceived risk and determine the best methodology for achieving high performance. This approach differs from other text encoding methods discussed earlier, as numerical data is converted into text format for the model to process using prompts. The prompts used in the following section are included in the Appendix \ref{app:prompts}, and follow the best practice from existing literature \cite{OpenAIPromptEngineering2024, lin2024graphenhancedlargelanguagemodels}. \citet{sui2024tablemeetsllmlarge} showed that contextual information about the tabular features enables a 0-shot framework to outperform 1-shot prompting methodology.

While LLMs perform very well in 0-shot settings \cite{kojima2023large}, there is significant evidence that shows that LLMs perform better in a k-shot setting \cite{clark2018think}; the ARC benchmark uses 25-shot prompts in the Eleuther AI evaluation harness \cite{eval-harness}. The problem with deploying a k-shot framework in this setting is that the SEC Filings are very long (13,267 tokens). Despite the increase in the context-window length of some newer LLMs, many new models are capped at 8192 tokens or below \cite{gpt-j, jiang2023mistral, touvron2023llama, Llama3modelcard}, and some other studies have shown performance deterioration as the input sequence increases \cite{li2024longcontext}. Fitting several examples of the task in the input sequence is impossible for many of the data samples, which means that k-shot performance is not reported for this task. 

We tested several models\footnote{\textit{gpt-3.5-turbo-0125}, \textit{gpt-4-turbo-2024-04-09} and \textit{gpt-4o-2024-05-13}, \textit{Llama-3 8B}} using the prompting structure laid out in Appendix \ref{app:prompts}. The models provide a good representation of the current state-of-the-art \cite{chiang2024chatbot}.

\subsubsection{LoRA Adaptation}
\label{sec:LoRA}

To adapt the LLMs we use LoRA (\textbf{Lo}w \textbf{R}ank \textbf{A}daptation) \cite{hu2021lora} fine-tuning. This technique involves optimizing the rank-decomposition matrices, $A$ \& $B$, of the change in model weights ($\Delta W$), where $W^{'}$ are the new model weights and $W$ are the pre-trained weights.
\begin{align}
    W^{'} &= W + \Delta W \\
    &= W + BA
\end{align}
The advantage is that it requires a lot less memory to fine-tune a model and in contrast to some parameter-efficient fine-tuning methods, adaptation can take place through the entire model stack.

\begin{table}[!b]
    \small
    \centering
    \begin{tabular}{c c c c c c c}
        \toprule
        \multirow{2}{*}{Data} & \multirow{2}{*}{Features} & \multirow{2}{*}{Av.} & \multicolumn{4}{c}{Quarters} \\
        \cline{4-7}
         & & & 1 & 2 & 3 & 4 \\
        \midrule
         \multirow{2}{*}{\( \mathbf{N} \)} & \({M + F + C} \) & \textbf{52.8} & \textbf{48.3} & \textbf{53.3} & \textbf{54.0} & \textbf{55.7} \\
         & \({C} \) & 44.7 & 41.9 & 43.6 & 46.5 & 46.7 \\
         \cline{2-7}
         \multirow{5}{*}{\( \mathbf{A} \)} & LM & 50.6 & 46.8 & 51.2 & 52.1 & 52.4 \\
         & LDA & 50.9 & 50.3 & 52.3 & 50.8 & 50.2 \\
         & HEC & \underline{\textbf{53.6}} & \underline{\textbf{50.7}} & \underline{\textbf{54.6}} & 54.1 & \underline{\textbf{56.0}} \\
         & \( E_{\text{DRoBERTa}} \) & 52.8 & 48.2 & 52.9 & 55.4 & 54.8 \\
         & MP-NET & 51.0 & 46.8 & 52.5 & \underline{\textbf{56.1}} & 48.4 \\
         \cline{2-7}
         \multirow{5}{*}{\( \mathbf{T} \)} & LM & 34.4 & 36.6 & 33.4 & 30.9 & 36.8 \\
         & LDA & 34.8 & 36.6 & 35.0 & 30.3 & \textbf{37.1} \\
         & HEC & \textbf{38.1} & \textbf{39.8} & \textbf{38.0} & 38.9 & 35.8 \\
         & \( E_{\text{DRoBERTa}} \) & 36.0 & 36.8 & 35.8 & 36.2 & 35.2 \\
         & MP-NET & 35.9 & 32.4 & 35.0 & \textbf{39.9} & 36.3 \\
         \bottomrule
    \end{tabular}
    \caption{The accuracy using the XGBoost model across different feature sets and text encoding methods. 
    \( \mathbf{N} \) refers to instances where only numeric information is used. \( \mathbf{T} \) refers to text-based data types. \( \mathbf{A} \) indicates all data types combined (\(M + F + C + T\)). \textbf{Bold}  indicates the best results for each of the data configurations; \underline{underline} indicates the best results across all configurations.}
    \label{tab: diff_models_rf_deep}
\end{table}

\section{Results}

\begin{table*}[t]
    \small
    \centering
    \begin{tabular}{c c c c c c c | c c c }
        \toprule
        \multirow{2}{*}{Model} & \multirow{2}{*}{Features} & \multirow{2}{*}{Average} & \multicolumn{4}{c}{Quarters} & \multicolumn{3}{c}{Feature Importances} \\
        \cmidrule{4-7}\cmidrule{8-10}
         & & & 1 & 2 & 3 & 4 & \( \mathbf{M} \) & \( \mathbf{F} \) & \( \mathbf{T} \) \\
        \midrule
        \multirow{5}{*}{XGBoost} 
        & \( \mathbf{N} \) & 52.8 & 48.3 & 53.3 & 54.0 & 55.7 & 0.659 & 0.341 & - \\
        & \( \mathbf{A}_{\text{HEC}} \) & \underline{\textbf{53.9}} & 50.7 & \underline{\textbf{54.6}} & \underline{\textbf{54.1}} & 56.0 & 0.117 & 0.188 & 0.695 \\
        & \( \mathbf{T}_{\text{HEC}} \) & 38.1 & 39.8 & 38.0 & 38.9 & 35.8 & - & - & - \\
        & \( \mathbf{N} + \mathbf{T}_{\text{GPT-4o Est.}} \) & 53.8 & 52.7 & 50.9 & 54.0 & \underline{\textbf{57.3}} & 0.588 & 0.302 & 0.101 \\
        & \( \mathbf{N}_{\text{XGB Est.}} + \mathbf{T} \) & 51.8 & \underline{\textbf{53.7}} & 51.2 & 50.8 & 51.3 & \multicolumn{2}{c}{0.573} & 0.427 \\
        \midrule
        \multirow{4}{*}{GPT-4o} 
        & \( \mathbf{N} \) & 31.4 & 33.7 & 31.6 & 31.1 & 29.3 & - & - & - \\
         & \( \mathbf{A} \) & 40.2 & 43.9 & 40.3 & 38.8 & 37.9 & - & - & - \\
         & \( \mathbf{T} \) & \textbf{49.6} & \textbf{49.3} & \textbf{52.2} & \textbf{52.4} & \textbf{44.6} & - & - & - \\
         & \( \mathbf{T} + \mathbf{N}_{\text{XGB Est.}} \) & 32.3 & 33.9 & 30.8 & 36.7 & 27.7 & - & - & - \\
         \midrule\midrule
        \multicolumn{2}{c}{XGBoost-\( \mathbf{N} \cup \text{GPT-4o-}\mathbf{T} \)} & 69.9 & 70.5 & 73.2 & 71.5 & 64.3 & - & - & - \\
        \bottomrule
    \end{tabular}
    \caption{Accuracy across different model and data configurations. The notation is consistent to Table \ref{tab: diff_models_rf_deep}. Feature importances for \( \mathbf{M}, \mathbf{F}, \mathbf{T} \) are impurity scores averaged across lags. \(\mathbf{N}_{\text{XGB Est.}}\) and \(\mathbf{T}_{\text{GPT-4o Est.}}\) represent the subscript model's estimate and implied internal probability of that estimate using the features represented by the bold letter, both are the probability and estimate are used as features.}
    \label{tab:overall}
\end{table*}

\begin{table}[t]
    \small
    \centering
    \begin{tabular}{c c c c c c c}
        \toprule
        \multirow{2}{*}{Data} & \multirow{2}{*}{Model} & \multirow{2}{*}{Av.} & \multicolumn{4}{c}{Quarters} \\
        \cline{4-7}
         & & & 1 & 2 & 3 & 4 \\
         \midrule
         \multirow{5}{*}{\( \mathbf{N} \)} & Llama & 32.3 & 35.1 & \textbf{35.8} & 29.0 & 29.3 \\
         & Llama-LoRA & \textbf{35.5} & \textbf{35.4} & 34.2 & \textbf{37.8} & 34.7 \\
         & GPT-3.5 & 32.6 & 32.9 & 33.4 & 31.9 & 32.3 \\
         & GPT-4 & 34.1 & 34.2 & 32.9 & 33.0 & \textbf{36.3} \\
         & GPT-4o & 31.4 & 33.7 & 31.6 & 31.1 & 29.3 \\
         \cline{2-7}
         \multirow{5}{*}{\( \mathbf{A} \)} & Llama & 35.5 & 35.4 & 35.8 & 37.0 & 33.9 \\
         & Llama-LoRA & 37.5 & 36.4 & 37.0 & 37.6 & 38.8 \\
         & GPT-3.5 & \textbf{44.5} & \textbf{49.0} & \textbf{42.4} & \textbf{46.0} & \textbf{40.6} \\
         & GPT-4 & 38.3 & 39.8 & 36.1 & 38.0 & 39.3 \\
         & GPT-4o & 40.2 & 43.9 & 40.3 & 38.8 & 37.9 \\
         \cline{2-7}
         \multirow{5}{*}{\( \mathbf{T} \)} & Llama & 35.6 & 35.4 & 36.1 & 37.0 & 33.9 \\
         & Llama-LoRA & 37.0 & 38.1 & 36.8 & 37.1 & 35.9 \\
         & GPT-3.5 & 46.4 & 47.3 & 47.5 & 45.5 & 45.2 \\
         & GPT-4 & 48.5 & 47.8 & 48.5 & 50.3 & \underline{\textbf{48.1}} \\
         & GPT-4o & \underline{\textbf{49.6}} & \underline{\textbf{49.3}} & \underline{\textbf{52.2}} & \underline{\textbf{52.4}} & 44.6 \\
        \bottomrule
    \end{tabular}
    \caption{Accuracy using the generative models. The notation is consistent to Table \ref{tab: diff_models_rf_deep}. All models are tested in 0-shot besides Llama-3 8B which is fine-tuned using LoRA.}
    \label{tab:gen_models}
\end{table}

The results from the XGBoost baseline are outlined in Table \ref{tab: diff_models_rf_deep}. It is clear that there is some information in the text since almost all text encoding methods perform above chance. However, none of the individual text feature sets outperform the numeric baselines, indicating that fundamental and macroeconomic variables are more critical for prediction. Combining features yields a performance boost, particularly when HEC features are integrated with numerical data.

Table \ref{tab:gen_models} highlights intriguing behavior in generative models. With a zero-shot prompt, performance using only numerical data is near random. Interestingly, GPT-class models perform better using text alone than with all data types, suggesting that numerical information may hinder their predictive accuracy. GPT-3.5, despite being older, achieves the best performance with all features. It also appears that LoRA enables better relative performance on numerical data - Llama-3 8B LoRA is the best-performing model on entirely numerical information and is the only model with no performance degradation when all features are considered as opposed to just text. Overall, generative models excel at decoding text data for this task.

Table \ref{tab:overall} takes the best-performing text features from the XGBoost framework, HEC, and provides a comparison to the best-performing generative model, GPT-4o. Interestingly, GPT-4o utilises the text alone much better than any of the encoder-based methods, but when all feature-types are considered the XGBoost-HEC configuration is the best-performing methodology. 

In addition, the models pick up on different signals. The final row of Table \ref{tab:overall} shows that the proportion of samples where at least one of the XGBoost-\( \mathbf{N} \) and GPT-4o-\( \mathbf{T} \) is correct (69.87) is significantly higher than any of the individual models. As a result, we provide comparisons where the estimate and class probability of the XGBoost-\( \mathbf{N} \) and GPT-4o-\( \mathbf{T} \) are included in the prompt or feature set. Both the estimate and the probability that each model assigns to the estimate are used in the prompt or as features. The combination methods all underperform the XGBoost-\( \mathbf{A}_{\text{HEC}} \) configuration, but this performance gap provides an opportunity for future research on ensembling methods.

\section{Interpretability}

One of the disadvantages of generative LLMs is that, to a large degree, they are black boxes. While some work has successfully used attention weights and internal model states to analyse generated prompts \cite{serrano-smith-2019-attention, wang2022interpretabilitywildcircuitindirect}, this is not a mature research area. Much of the mechanistic interpretability literature has focused on toy models \cite{elhage2021transformer}, and while there has been some progress made on extracting features from larger models using sparse autoencoders \cite{templeton2024scaling}, the field is still very far from completely solving interpretability within LLMs. In the absence of a complete solution, it is worth acknowledging the interpretable features that traditional methods use. 
Regulation in major economies increasingly emphasizes explainability alongside performance \cite{EuropeanCommission2020, USCongress2022, UKSecretaryOfState2022}. The XGBoost-\( \mathbf{A}_{\text{HEC}}\) framework not only achieves the best performance among the models in this paper but also enables users to interpret its decisions. This section provides an example of how we can use this framework to understand the reasons behind decisions.

\begin{figure}[ht]
    \centering
    \begin{subfigure}[h]{\linewidth}
        \centering
        \includegraphics[width=\linewidth]{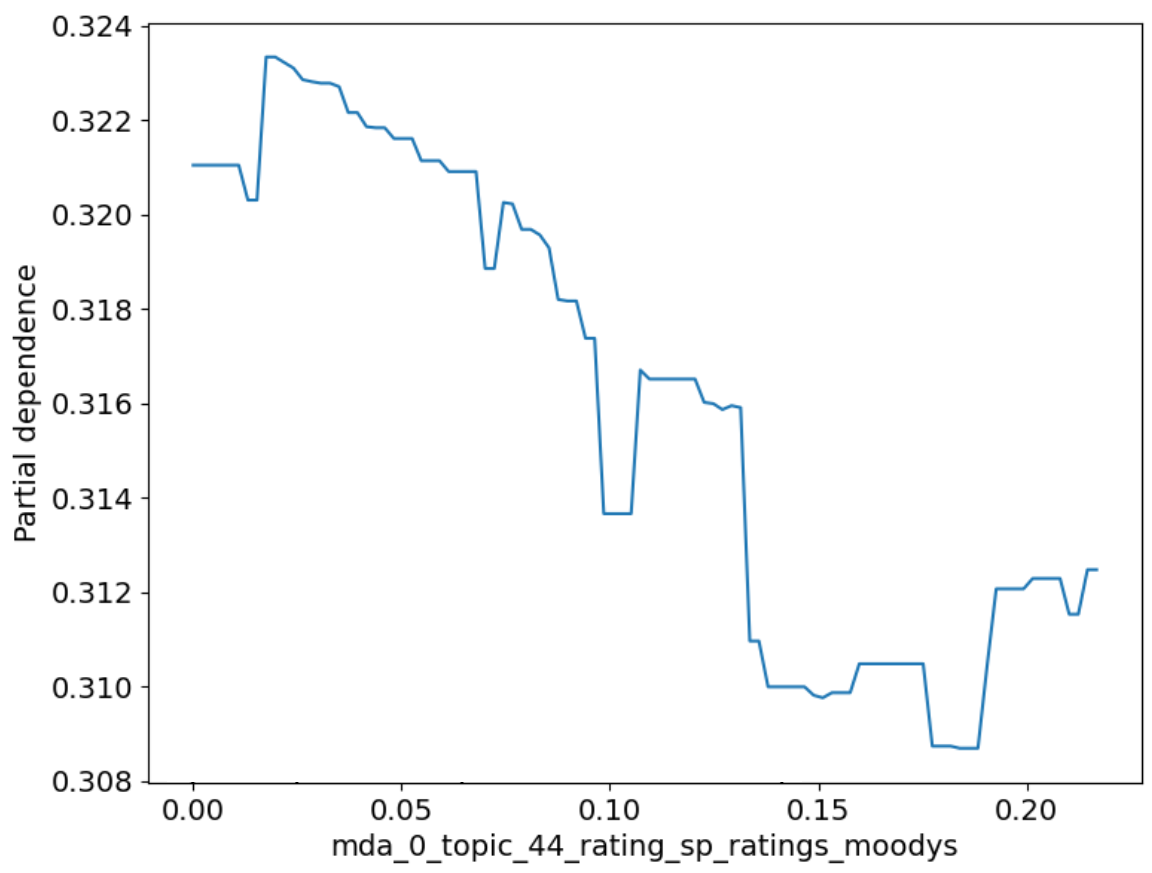}
        \caption{PDP of "rating\_sp\_ratings\_moodys" cluster \& "Up" class.}
        \label{fig:ratings_subfig}
    \end{subfigure}
    \begin{subfigure}[h]{\linewidth}
        \centering
        \includegraphics[width=\linewidth]{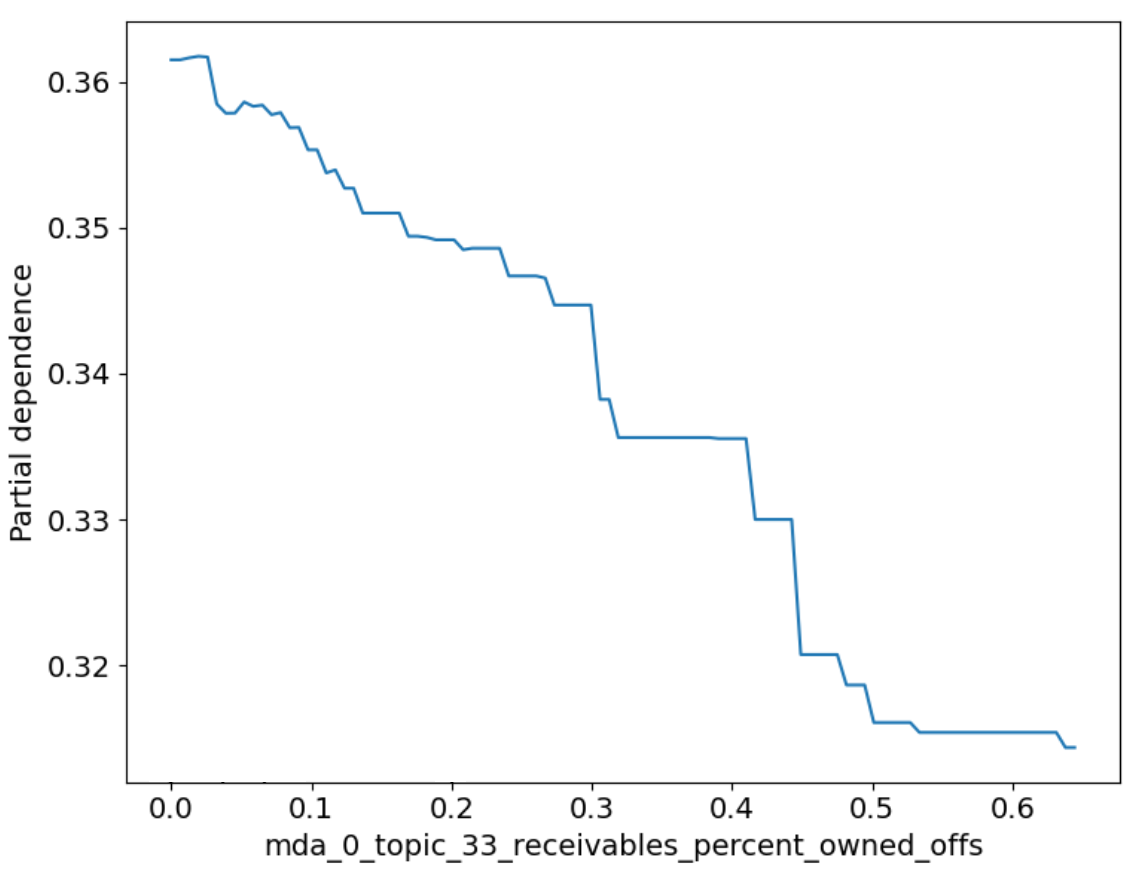}
        \caption{PDP of "receivables\_percent\_owned\_offs" cluster \& "Down" class.}
        \label{fig:receivables_subfig}
    \end{subfigure}
    \caption{Partial Dependence Plots (PDP) of text-based features against different target classes.}
    \label{fig:pdp}
\end{figure}

The most obvious advantage of a feature-based system is that important features can be identified. Table \ref{tab:overall} provides an example of feature importance that can be used to infer the modality preference of model configurations. It is possible to conduct even more granular feature analysis by looking at the contribution of individual features. Figure \ref{fig:pdp} shows the partial dependence plot of some of the individual text features on the "Up" and "Down" classes. From the plot in Figure \ref{fig:ratings_subfig} we can infer that as ratings are discussed more in a company's filing, there is a reduced chance of the the credit rating being upgraded. We can also infer from Figure \ref{fig:receivables_subfig} that as companies talk about receivables - the money owed to the company - there is a reduced chance of the credit rating being downgraded. Both provide valuable insights and are examples of how a traditional feature-based methodology can be leveraged for increased interpretability.

\section{Conclusion}

The paper shows that while LLMs are good at encoding textual data and inferring signals that traditional methods cannot pick up on when combined with numerical data in the prompt there is performance deterioration. The other advantage is that traditional methods offer increased interpretability and a better understanding of the mechanisms behind certain predictions. In addition, traditional approaches don't suffer to the same extent from complications associated with training data contamination and memorization \cite{ozdayi-etal-2023-controlling, lu2024scalinglawsfactmemorization} since the models used for the traditional features are much smaller than the generative models and memorisation in LLMs exhibits scaling law behaviour. While it is not impossible that the text data used in this paper was included in the training of the studied LLMs, any potential influence would likely have a greater impact on generative models, thereby reinforcing the findings of this paper.

There has been some work jointly encoding text using generative LLMs with time-series information \cite{liu2024unitime}, but more work needs to be done to determine the best methodology for combining long text sequences with numerical information while utilizing the benefits of generative LLM natural language understanding. This paper shows that combining multimodal information within the prompt is not sufficient.

\section{Limitations}

The task above uses a balanced dataset, which is good for testing the different methodologies' ability to discern the signals that are predictive of a rise or fall in credit ratings, but is poor for assessing how good the models would be in a real-world context where almost all of the ratings stay the same. Despite the data being taken from across all US equities over a 23-year time period, the balanced dataset is relatively small, with only 3441 samples for the Lag 1 configuration and 2142 samples for the Lag 4 configuration. There are plenty of prominent datasets that are smaller, but the size reduces the scope for complex and specialized models to be deployed on this task in favour of more robust, simple models. 

Another limitation is that the text used in this paper is produced by the companies themselves, who the goal of conveying a positive viewpoint to investors. More objective publication venues may produce different insights about the future direction of a company.

We also assume that the credit rating methodology remains the same between the train and test sets. This is an assumption that is made by the rest of the literature, and our training set is spread over an 18 year period, however it does not rule out the possibility that the results are only valid over the time period that was tested. Due to the size of the dataset we were restricted from using a masked temporal cross-validation evaluation framework, which would have left insufficient data for training for some years.

The LoRA fine-tuning methodology outlined in this paper is a parameter-efficient technique and has been shown to be competitive in a variety of settings \cite{hu2021lora}, but can be outperformed in some tasks by full fine-tuning and other adapter-based methods \cite{xu2023parameterefficientfinetuningmethodspretrained}. We compared the performance of the LoRA implementation in this paper to that of QLoRA \cite{dettmers2023qlora}, which produced marginally worse results. However, it is possible that other fine-tuning techniques would have produced better results. Further to this, it is possible that if a better model than Llama-3 8B had been fine-tuned we would have seen even better results from the generative LLMs. The computational constraints placed on us are not dissimilar to those that other researchers face, which makes the results in this paper valid while perhaps not exhaustive.

\bibliography{acl_latex}
\appendix

\section{Greedy Decoding}
\label{app:greedy_decode}

Generative models can produce unpredictable outputs \cite{fadaee-monz-2020-unreasonable, stureborg2024large}, which necessitates the use of constrained generation when an LLM forms part of a larger architecture. For the purposes of this work, we use a greedy search to infer the probability of one of the following labels appearing next in the sequence: "up", "down" or "same". 

\section{Dataset Size}
\label{app:dataset_size}

The size of each of the dataset splits are outlined in Table \ref{tab:dataset_size}.

\begin{table}[h]
     \centering
    \begin{tabular}{c|ccc}
        \# Quarters & Train & Val & Test \\
        \hline
        1 & 2,642 & 389 & 410 \\
        2 & 1,748 & 374 & 377 \\
        3 & 1,595 & 351 & 376 \\
        4 & 1,445 & 325 & 372 \\
    \end{tabular}
    \caption{Dataset sizes}
    \label{tab:dataset_size}
\end{table}

\section{Fundamental Data}
\label{app:var_desc} 

The fundamental variables considered are outlined in Table \ref{tab:variables}.

\begin{table}[h]
    \centering
    \begin{tabular}{|c|c|l|}
        \hline
        \textbf{Variable} & \textbf{Type} & \textbf{Description} \\
        \hline
        \texttt{niq} & Float & Net Income (Loss) \\
        \texttt{ltq} & Float & Liabilities - Total \\
        \texttt{piq} & Float & Pretax Income \\
        \texttt{atq} & Float & Assets - Total \\
        \texttt{ggroup} & Char & GIC Groups \\
        \texttt{gind} & Char & GIC Industries \\
        \texttt{gsector} & Char & GIC Sectors \\
        \texttt{gsubind} & Char & GIC Sub-Industries \\
        \hline
    \end{tabular}
    \caption{Description of Variables}
    \label{tab:variables}
\end{table}

\section{Neural Network Implementations}
\label{app:hier_cr}

We also tested some more complex neural network (NN) approaches, which had underwhelming results. The \textbf{Hier}archical \textbf{C}redit \textbf{R}ating (HierCR) model is a framework that models the filings hierarchically. The challenge with using LLMs to encode the filings is the limited context window of encoder-based LLMs. There have been many different solutions to this problem, including sparse attention mechanisms \cite{beltagy2020longformer}, chunking \cite{sawhney-etal-2020-deep}, and feature-based extraction like the methods above \cite{loughran2011liability, drinkall-etal-2022-forecasting}. Our NN solution to this problem is to split the filing up into sentences and pass the sentence embeddings through an \textit{all-mpnet-base-v2} encoder to produce embeddings for the textual data. The text encoder replicates the structure in \cite{sawhney-etal-2020-deep}, the only material difference is that filing sentences substitute the social media posts in the first layer of the text-encoder. The text, macro and fundamental vectors across the previous quarter(s) are combined using a GRU layer \cite{chung2014empiricalevaluationgatedrecurrent}, the outputs are then passed through an attention layer to create a representation for each data type. These representations are combined using a bilinear transformation, which is passed through 3 linear layers followed by a ReLU \cite{agarap2019deeplearningusingrectified} activation function. Dropout is applied in the final linear-layer classification module.

 \begin{table}[h]
    \small
    \centering
    \begin{tabular}{c c c c c c c}
        \toprule
        \multirow{2}{*}{Model} & \multirow{2}{*}{Av.} & \multicolumn{4}{c}{Quarter} \\
        \cmidrule{3-6}
         & & 1 & 2 & 3 & 4 \\
        \midrule
        SA-LF & 40.62 & 39.87 & 40.05 & 41.17 & 41.39 \\
        SA-EF & 43.41 & 41.24 & 40.97 & 45.31 & 46.11 \\
        HierCR & 35.02 & 33.58 & 32.89 & 35.62 & 37.98 \\
        \bottomrule
    \end{tabular}
    \caption{Accuracy for more complex NN approaches using all data types (M+F+T).}
    \label{tab:emotion_lex_performance}
\end{table}

The other two architectures are \textbf{S}hared \textbf{A}ttention \textbf{L}ate \textbf{F}usion (SA-LF) and \textbf{S}hared \textbf{A}ttention \textbf{E}arly \textbf{F}usion (SA-EF). Both architectures only consider the first 512 tokens of each filing. The difference between the two architectures is when the attention layer is applied. For the SA-EF the model attends to all feature types together, whereas the SA-LF only combines the representations after the attention layer is applied to the individual data types. The final representation is the passed through the same linear-layer classification module as the HierCR. For all of the architectures above, we trained the model for 200 epochs with a patience value of 20 epochs.

The results from these models are poor in comparison to the more simple XGBoost models. This could be due to the size of the dataset, which does not provide the model enough data to train on without overfitting the training data. Complex models with many parameters require more data to fit properly. Given that the dataset is the largest balanced and complete dataset possible to make using US data, and that the size of the dataset considered in this paper is representative of a large number of other tasks, the results from this paper represent a significant contribution for dealing with problems of this nature.

\section{Prompts}
\label{app:prompts}

To encode the numerical and textual information into text form, we used the prompting structure outlined below. When the ablation study was carried out the prompt and data included was adjusted accordingly.\\\\
\textit{\textbf{\#\#\# System}:
You are trying to work out whether a company's credit rating is likely to go up, down, or stay the same given its recent credit ratings. Predict the likely movement in a company's credit rating for the next quarter, using historical credit ratings, quantitative financial data and macroeconomic data. The numeric data has been normalized and appears in order with the most recent first.\\
\textbf{\#\#\# Credit Rating Explanation:}\\
Credit ratings use the following scale, in order of increasing risk: 'AAA', 'AA+', 'AA', 'AA-', 'A+', 'A', 'A-', 'BBB+', 'BBB', 'BBB-', 'BB+', 'BB', 'BB-', 'B+', 'B', 'B-', 'CCC', 'CCC-', 'CC', 'C', 'SD'\\
\textbf{\#\#\# Fundamental Financial Indicators Defined:}\\
... \\
\textbf{\#\#\# Macroeconomic Variables Defined:}\\
... \\
\textbf{\#\#\# User:}\\
Your task is to classify the company into one of the following classes: "down", "same", "up".   
"down" means that you think the credit rating will go down in the next quarter, meaning the company is perceived as more risky.
"same" means that you think the credit rating will stay the same in the next quarter. 
"up" means that you think the credit rating will go up in the next quarter, meaning the company is perceived as less risky.
Please respond with a single label that you think fits the company best. \\ 
\vspace{0.2cm}
Classify the following numerical data:"""}

\subsection{Credit Rating Ranking}
\label{app:ranking}

One potential problem with the prompt outlined in Appendix \ref{app:prompts} is that the LLM may find it hard to correctly understand the ranking structure of credit ratings, which would limit the ability of an LLM to perform well on this task. To probe the LLMs ability to understand the relative rank of credit ratings we created the following prompt:\\\\
\textit{"""Two credit ratings will be given, the task is to determine which is higher on the following scale, which is ordered in descending order: \\\\
'AAA', 'AA+', 'AA', 'AA-', 'A+', 'A', 'A-', 'BBB+', 'BBB', 'BBB-', 'BB+', 'BB', 'BB-', 'B+', 'B', 'B-', 'CCC', 'CCC-', 'CC', 'C', 'SD'. \\\\
Please answer with the higher rating e.g. AAA vs. SD Answer: AAA.\\
<<rating\_X>> vs. <<rating\_Y>> Answer:"""\\\\}

The performance on this task across all rating combinations when prompting GPT-4o was 99.52\%. The only mistake was between C and CC. This high performance displays a very good understanding of the credit rating scale and justifies the setup of our prompt.

\section{S\&P Credit Rating Definitions}
\label{app:CR_definitions}

S\&P's definitions for each of the credit rating categories are outlined in Table \ref{tab:ratings}.

\begin{table*}[]
    \centering
    \begin{tabular}{|>{\raggedright\arraybackslash}m{2cm}|>{\raggedright\arraybackslash}m{14cm}|}
    \hline
    \textbf{Category} & \textbf{Definition} \\
    \hline
    AAA & An obligation rated 'AAA' has the highest rating assigned by S\&P Global Ratings. The obligor's capacity to meet its financial commitment on the obligation is extremely strong. \\
    \hline
    AA & An obligation rated 'AA' differs from the highest-rated obligations only to a small degree. The obligor's capacity to meet its financial commitment on the obligation is very strong. \\
    \hline
    A & An obligation rated 'A' is somewhat more susceptible to the adverse effects of changes in circumstances and economic conditions than obligations in higher-rated categories. However, the obligor's capacity to meet its financial commitment on the obligation is still strong. \\
    \hline
    BBB & An obligation rated 'BBB' exhibits adequate protection parameters. However, adverse conditions or changing circumstances are likely to lead to a weakened capacity of the obligor to meet its financial commitment on the obligation. \\
    \hline
    BB; B; CCC; CC; and C & Obligations rated 'BB', 'B', 'CCC', 'CC', and 'C' are regarded as having significant speculative characteristics. 'BB' indicates the least degree of speculation and 'C' the highest. While such obligations will likely have some quality and protective characteristics, these may be outweighed by large uncertainties or major exposures to adverse conditions. \\
    \hline
    BB & An obligation rated 'BB' is less vulnerable to nonpayment than other speculative issues. However, it faces major uncertainties or exposure to adverse business, financial, or economic conditions which could lead to the obligor's inadequate capacity to meet its financial commitment on the obligation. \\
    \hline
    B & An obligation rated 'B' is more vulnerable to nonpayment than obligations rated 'BB', but the obligor currently has the capacity to meet its financial commitment on the obligation. Adverse business, financial, or economic conditions will likely impair the obligor's capacity or willingness to meet its financial commitment on the obligation. \\
    \hline
    CCC & An obligation rated 'CCC' is currently vulnerable to nonpayment, and is dependent upon favorable business, financial, and economic conditions for the obligor to meet its financial commitment on the obligation. In the event of adverse business, financial, or economic conditions, the obligor is not likely to have the capacity to meet its financial commitment on the obligation. \\
    \hline
    CC & An obligation rated 'CC' is currently highly vulnerable to nonpayment. The 'CC' rating is used when a default has not yet occurred, but S\&P Global Ratings expects default to be a virtual certainty, regardless of the anticipated time to default. \\
    \hline
    C & An obligation rated 'C' is currently highly vulnerable to nonpayment, and the obligation is expected to have lower relative seniority or lower ultimate recovery compared to obligations that are rated higher. \\
    \hline
    SD & An obligation rated 'SD' is in default or in breach of an imputed promise. For non-hybrid capital instruments, the 'SD' rating category is used when payments on an obligation are not made on the date due, unless S\&P Global Ratings believes that such payments will be made within five business days in the absence of a stated grace period or within the earlier of the stated grace period or 30 calendar days.  \\
    \hline
    NR & This indicates that no rating has been requested, or that there is insufficient information on which to base a rating, or that S\&P Global Ratings does not rate a particular obligation as a matter of policy. \\
    \hline
    \end{tabular}
    \caption{S\&P Global Ratings Definitions \cite{spglobalratings}}
    \label{tab:ratings}
\end{table*}

\end{document}